\documentclass{jfm}
\usepackage[T2A]{fontenc} 
\usepackage[utf8]{inputenc} 
\usepackage[russian,english]{babel} 
\usepackage{graphicx}
\usepackage{epstopdf, epsfig}
\usepackage{color}
\usepackage{ulem}
\newcommand{\ba}{\begin{array}}
\newcommand{\ea}{\end{array}}
\newcommand{\be}{\begin{equation}}
\newcommand{\ee}{\end{equation}}
\newcommand{\bea}{\begin{eqnarray}}
\newcommand{\eea}{\end{eqnarray}}
\newcommand{\bfig}{\begin{figure}}
\newcommand{\efig}{\end{figure}}
\newcommand{\Bl}{\Bigl}
\newcommand{\Br}{\Bigr}

\newcommand{\pl}{\partial}
\newcommand{\dd}{{\rm d}}

\newcommand{\vp}{\varphi}
\newcommand{\vep}{\varepsilon}

\newcommand{\din}{\displaystyle\int\limits}

\newcommand{\dfrac}{\displaystyle\frac}
\shorttitle{Reflectionless wave propagation}
\shortauthor{S. M. Churilov and Y. A. Stepanyants}

\title{Reflectionless wave propagation on shallow water with variable bathymetry and current. II}

\author{Semyon M. Churilov\aff{1}
     \and Yury A. Stepanyants\aff{2,}\aff{3}
\corresp{\email{Yury.Stepanyants@usq.edu.au}}}

\affiliation{\aff{1} Institute of Solar-Terrestrial Physics of the Siberian Branch of the Russian Academy of Sciences, PO Box 291, Irkutsk, 664033, Russia
\aff{2} Scool of Sciences, University of Southern Queensland, West St., Toowoomba, QLD, 4350, Australia \\
\aff{3} Department of Applied Mathematics, Nizhny Novgorod State Technical University \\n.a. R. E. Alekseev, 24 Minin St., Nizhny Novgorod, 603950, Russia}

\begin{document}

\maketitle

\begin{abstract}
We show that in the linear approximation there are three classes of reflectionless wave propagation on a surface of shallow water in the channel with spatially varying depth, width, and current speed. 
Two of these classes have been described in our previous paper \citep{ChSt22}, and the third one was discovered recently and is described here. 
The general analysis of the problem shows that within the approach used in both our papers, these three classes, apparently, exhaust all possible cases of exact solutions of the problem considered.  
We show that the reflectionless flow can be global at certain conditions, i.e. it can exist on the entire $x$-axis.
There are also reflectionless flows which exist only on the limited intervals of the $x$-axis. 
\end{abstract}


\section{Introduction}
\label{Sect01}

The conditions under which waves propagate in inhomogeneous media without reflection and scattering are of significant physical interest and are important for practical purposes. 
Under such conditions, wave energy can be the most effectively transmitted over long distances. 
In the recent paper \citep{ChSt22} we found two classes of shallow water flows in channels with the variable width $W(x)$ and bottom profile $z_B = B(x)$ (see Fig.~1) which provide reflectionless (RL) propagation of long surface waves. 
However, the profiles found in that paper do not exhaust all possible classes of flows that provide RL wave propagation.
Recently we found one more class of such flows which is noticeably different from those that have been already studied. 
The aim of this paper is to describe this specific class of flows and complete the description of all possible classes of RL flows, at least within the framework of the approach based on a factorization of wave equations (see below).
Thus, this paper can be considered as the continuation of our previous paper \citep{ChSt22}.
\begin{figure}
\centerline{\includegraphics[width=0.8\textwidth]{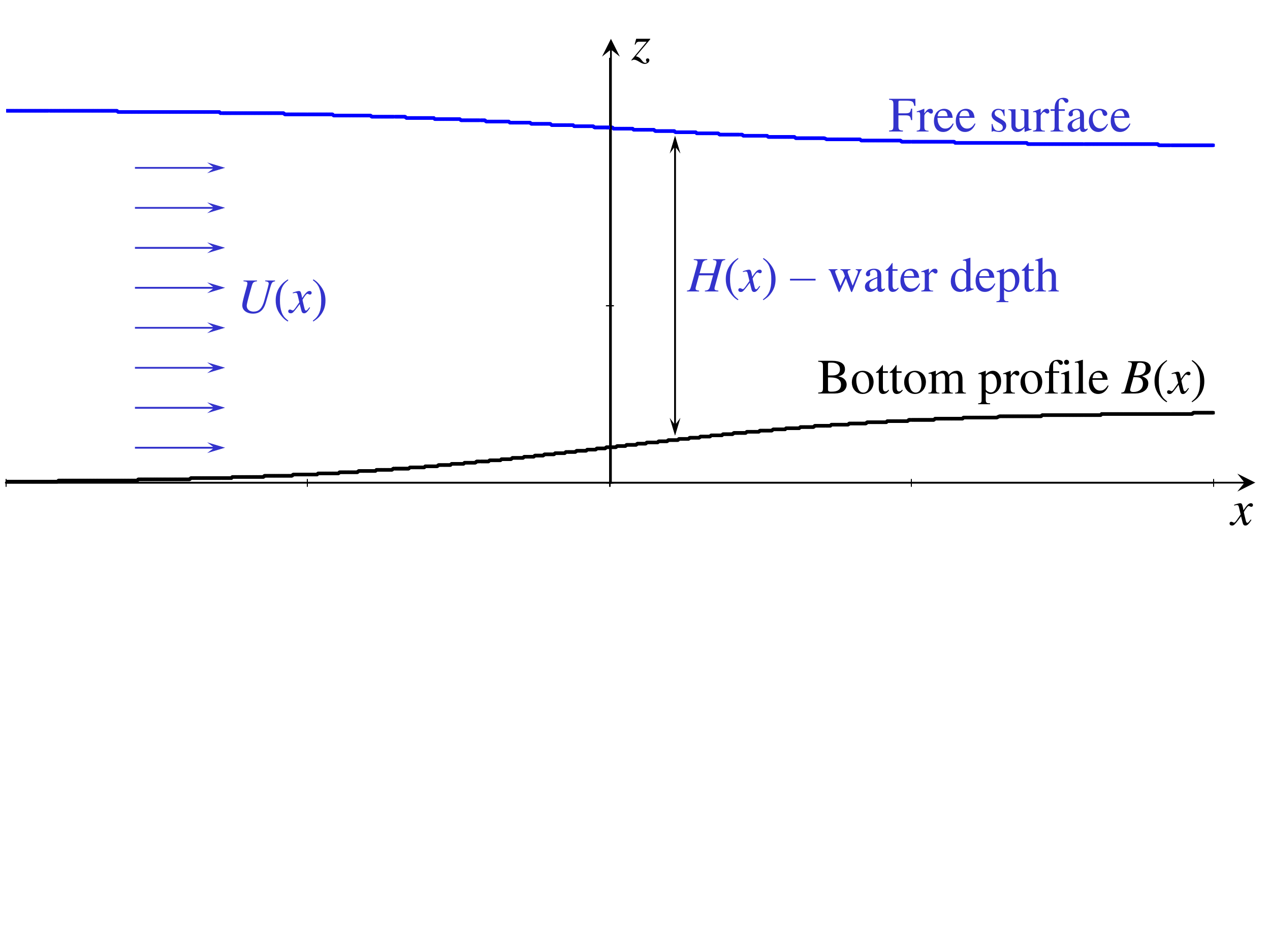}}%
\vspace{-3.5cm}
\caption{Sketch of the flow configuration in the vertical plane.}
\label{f00}
\end{figure}
\section{Basic equations}
\label{Sect02}

In a stationary flow, the mass conservation,
 \be
 U(x)H(x)W(x) = {\rm const},
 \label{Flux}
 \ee
together with the Bernoulli law,
 \be
 \frac{1}{2}\,U^2(x) + g\Bl(H(x) + B(x)\Br) = {\rm const},
 \label{Bern}
 \ee
provide independent variation along the longitudinal channel $x$-axis the flow $U(x)$ and wave $c(x)=\sqrt{gH(x)}$ speeds (we assume that both of them are positive).
 Here $H(x)$ is the water depth in the channel, and $g$ is the acceleration due to gravity.
 
In the shallow-water approximation, the linearised Euler equation,
 \be
 \dfrac{\pl \tilde u}{\pl t} + \dfrac{\pl(U\tilde u)}{\pl x} = -g\dfrac{\pl\eta}{\pl x},
 \label{Euler}
 \ee
where $\tilde u(x,t)$ is the velocity perturbation and $\eta(x,t)$ is the deviation of water surface from the equilibrium, and mass balance equation,
 \be
 \dfrac{\pl\eta}{\pl t} + \dfrac{1}{W}\,\dfrac{\pl}{\pl x}\Bl[W(U\,\eta +
 H\,\tilde u)\Br] \equiv
 \dfrac{\pl\eta}{\pl t} + HU\dfrac{\pl}{\pl x}\left(\dfrac{\eta}{H} +
 \dfrac{\tilde u}{U}\right) = 0,
 \label{eta}
 \ee
can be reduced to a single equation by two ways.

On the one hand, we can introduce the velocity potential $\vp(x)$ by setting $u(x,t) = \pl\vp(x,t)/\pl x$, express $\eta$ in terms of $\vp(x)$ using (\ref {Euler}), and substitute in equation (\ref{eta}). 
Then, we can present $\vp$ in the form $\vp(x, t) = a(x)\psi(x,t)$ and obtain equation for $\psi$ in two equivalent forms (see  \citep{ChSt22}):
 \be
 \ba{l}
 \left\{\dfrac{\pl}{\pl t} + \Bl[U(x) - c(x)\Br]\,\dfrac{\pl}{\pl x} +
 U(x)\left(\dfrac{U'(x)}{U(x)} - \dfrac{c'(x)}{c(x)}\right)\right\}
 \left(\dfrac{\pl}{\pl t}+ \Bl[U(x)+c(x)\Br]\dfrac{\pl}{\pl x}\right)\psi
 \equiv  \\ \\
 \left\{\dfrac{\pl}{\pl t} + \Bl[U(x) + c(x)\Br]\,\dfrac{\pl}{\pl x} +
 U(x)\left(\dfrac{U'(x)}{U(x)} - \dfrac{c'(x)}{c(x)}\right)\right\}
 \left(\dfrac{\pl}{\pl t}+ \Bl[U(x)-c(x)\Br]\dfrac{\pl}{\pl x}\right)\psi
 = 0,
 \ea
 \label{WEq2}
 \ee
(hereinafter the prime denotes the $x$-derivative) provided that the following equations are fulfilled:
 \be
 \dfrac{\dd a}{\dd x} = \dfrac{{\cal B}\,c^2U}{c^2-U^2} \quad \mbox{and}
 \quad a(x) = \Bl[c(x)\,U(x)\Br]^{1/2},
 \label{Eq-a1}
 \ee
or, equivalently,
 \be
 \dfrac{\dd(c\,U)}{\dd x} \equiv c(x)\,\dfrac{\dd U}{\dd x} + U(x)\,
 \dfrac{\dd c}{\dd x} = \dfrac{{\cal B}\,c^{5/2}(x)\,U^{3/2}(x)}{c^2(x)-U^2(x)}\,.
 \label{Eq-aa}
 \ee
 Here ${\cal B} = {\rm const}$ (in \citep{ChSt22} it was denoted by $D$), and $c^{1/2}(x)$ and $U^{1/2}(x)$ should be considered as positive functions. The general solution of (\ref{WEq2}) is the sum of two independent waves of arbitrary shape traveling with different speeds,
 \be
 \psi(x,t) = \psi_1\left(t-\int\dfrac{\dd x}{U(x)+c(x)}\right) +
 \psi_2\left(t-\int\dfrac{\dd x}{U(x)-c(x)}\right).
 \label{psi}
 \ee
 It is the equation (\ref{Eq-aa}) which specifies such a relationship between $U(x)$ and $c(x)$ that the inhomogeneous flow becomes reflectonless.

On the other hand, we can introduce another potential, $\phi$, by setting $W(x)\eta(x,t) = \pl\phi(x,t)/\pl x$, and then, by integrating (\ref{eta}), express $\tilde u$ in terms of $\phi$. Substituting $\tilde u(\phi)$ into  (\ref{Euler}), we get the equation:
 \[
 \left(\dfrac{\pl}{\pl t}+U\dfrac{\pl}{\pl x}+2U'\right)
 \left(\dfrac{\pl\phi}{\pl t}+U\dfrac{\pl\phi}{\pl x}\right)=
 c^2W\dfrac{\pl}{\pl x}\left(\dfrac{1}{W}\,\dfrac{\pl\phi}{\pl x}\right).
 \]
 Note that this equation is valid in the absence of flow as well. Now, we put $\phi(x,t) = A(x)\chi(x,t)$, where $A(x) > 0$, eliminate $W$ using  (\ref{Flux}), and obtain
 \be
 \ba{l}
 \dfrac{\pl^2\chi}{\pl t^2} + (U^2-c^2)\,\dfrac{\pl^2\chi}{\pl x^2} +
 2U\,\dfrac{\pl^2\chi}{\pl t\pl x} +
 2\left(U\,\dfrac{A'}{A} + U'\right)\dfrac{\pl\chi}{\pl t}
   \\ \\ \phantom{wwa}
 +\,\left[2(U^2-c^2)\,\dfrac{A'}{A} - c^2\,\dfrac{U'}{U} + 3UU' - 2cc'
 \right]\dfrac{\pl\chi}{\pl x} + T(x)\chi = 0,
 \ea
 \label{WEq-z}
 \ee
 where $T(x)$ is defined through the equation:
 \[
 A(x)\,T(x) = (U^2-c^2)A'' + \Bl[3UU' - c^2(\ln U)' - 2cc'\Br]A'.
 \]
It can be readily shown that $T(x) \equiv 0$ if $A(x)$ obeys the equation:
 \be
 \dfrac{\dd A}{\dd x} = \dfrac{{\cal C}}{U(x)\Bl[U^2(x) - c^2(x)\Br]}\,,
 \qquad {\cal C} = \mbox{const}.
 \label{Eq-A}
 \ee

Following \citep{ChSt22}, consider the model equation:
 \be
 \left(\dfrac{\pl}{\pl t} + v_1(x)\,\dfrac{\pl}{\pl x} + {\cal F}(x)\right)
 \left(\dfrac{\pl}{\pl t} + v_2(x)\,\dfrac{\pl}{\pl x}\right){\cal H}(x,t) = 0.
 \label{ModEq}
 \ee
At least one of its solutions has the form of a traveling wave,
 \[
 {\cal H}(x,t) = {\cal H}_1\left(t - \int\dfrac{\dd x}{v_2(x)}\right),
 \]
 where ${\cal H}_1(z)$ is an arbitrary function. Let us remove brackets in (\ref{ModEq}):
 $$
 \dfrac{\pl^2 \cal H}{\pl t^2} + v_1(x)v_2(x)\,\dfrac{\pl^2 \cal H}{\pl x^2} + \Bl[v_1(x) + v_2(x)\Br]\dfrac{\pl^2 \cal H}{\pl t\pl x} $$
 \be
{} + {\cal F}(x)\,\dfrac{\pl {\cal H}}{\pl t} + \Bl[v_1(x)v'_2(x) + {\cal F}(x)v_2(x)\Br] \dfrac{\pl {\cal H}}{\pl x} = 0,
 \label{ModEq1}
 \ee
and find conditions when this equation coincides with equation (\ref{WEq-z}) provided that $T(x)\equiv 0$:
\refstepcounter{equation}
$$
 v_1(x)v_2(x) = U^2(x) - c^2(x), \quad 
 v_1(x) + v_2(x) = 2U(x)
 ,  
 \eqno{(\theequation{\mathit{a},\mathit{b}})}
 \label{R1}
$$
\be
 {\cal F} = 2\left(U\,\dfrac{A'}{A} + U'\right), \qquad
 v_1v'_2  + {\cal F} v_2 = 2(U^2-c^2)\,\dfrac{A'}{A} - c^2\,\dfrac{U'}{U}+3UU'-2cc'.  
 \label{R2}
 \ee
 Equations (\ref{R1}$\,a,b$) are fulfilled if either $v_1 = U - c$,
 $v_2 = U + c$, or $v_1 = U + c$, $v_2 = U - c$. In both these cases  (\ref{R2}) yields, up to an unimportant numerical factor, 
 \be
 A(x) = \Bl[c(x)U(x)\Br]^{-1/2} \equiv a^{-1}(x).
 \label{A}
 \ee
With this in mind, (\ref{Eq-A}) can be written as
 \be
 \dfrac{\dd A^{-1}}{\dd x} \equiv \dfrac{\dd a}{\dd x} =
 \dfrac{{\cal C}\,c(x)}{c^2(x) - U^2(x)}\,, \quad \mbox{or} \quad
 \dfrac{\dd(c\,U)}{\dd x} =
 \dfrac{{\cal C}\,c^{3/2}(x)\,U^{1/2}(x)}{c^2(x) - U^2(x)}\,.
 \label{Eq-AA}
 \ee
When  (\ref{Eq-AA}) is fulfilled, the function $\chi(x, t)$ obeys the same equation (\ref{WEq2}) as $\psi(x, t)$, and in the general case is also equal to the sum of two traveling waves of arbitrary shape,
 \be
 \chi(x,t) = \chi_1\left(t-\int\dfrac{\dd x}{U(x)+c(x)}\right) +
 \chi_2\left(t-\int\dfrac{\dd x}{U(x)-c(x)}\right).
 \label{chi}
 \ee
 However, RL  propagation of these two waves is now secured by  (\ref{Eq-AA}), which differs from  (\ref{Eq-aa}). The physical variables $\tilde u$ and $\eta$ are related to $\phi$ and $\chi$ in the following way:
 \be
 \ba{l}
 \tilde u(x,t) = -\dfrac{1}{H\,W}\left[\dfrac{\pl\phi}{\pl t} +
 U\,\dfrac{\pl\phi}{\pl x}\right] = -\dfrac{1}{H\,W\,a}
 \left[\dfrac{\pl\chi}{\pl t} + U\,\dfrac{\pl\chi}{\pl x} -
 \dfrac{a'}{a}\,U\,\chi(x,t)\right],
   \\ \\
 \eta(x,t) = \dfrac{1}{W}\,\dfrac{\pl\phi}{\pl x} = \dfrac{1}{W\,a}
 \left[\dfrac{\pl\chi}{\pl x} - \dfrac{a'}{a}\,\chi(x,t)\right].
 \ea
 \label{uz-chi}
 \ee

The problem of finding RL  profiles $c(x)$ and $U(x)$ has an infinite number of solutions, since these two velocities are related by only one equation (\ref{Eq-aa}) or (\ref{Eq-AA}).
To find specific solutions, it is necessary to set additionally either one of the speeds, or the relationship between them.
Below we will assume that $c(x)$ is known. 
Other options (setting $U(x)$ or the functional relationship between the speeds) have been considered in the context of equation (\ref{Eq-aa}) by \citet{ChSt22}.

At first glance, the differences between  (\ref{Eq-aa}) and (\ref{Eq-AA}) are insignificant, but the properties of the RL velocity profiles that satisfy them notably differ.
Only when ${\cal B} = {\cal C} = 0$, both equations lead to the same relation between the velocities, $a(x) = \mbox{const}$, or
 \be
 c(x)\,U(x) = \Pi = \mbox{const} > 0.
 \label{M1}
 \ee
In this class of RL flows (let's call it class A) the following relations are fulfilled owing to equations (\ref{M1}) and (\ref{Flux}):
\refstepcounter{equation}
$$
 W(x)\,H^{1/2}(x) = \mbox{const} \quad \mbox{and} \quad
 U(x)\,H^{1/2}(x) = \mbox{const},
 \eqno{(\theequation{\mathit{a},\mathit{b}})}
 \label{scc}
 $$
 so that $U(x)/W(x) = \mbox{const}$, i.e. the wider the channel, the higher the fluid velocity.

In studies of RL propagation of long surface waves in channels without current, the relation (\ref{scc}\,$b$) plays an important role (see, for example, \citep{Pelin-17, Pelinovsky-17}).
It distinguishes the so-called self-consistent channels -- the only class of channels with regular $W (x)$ and $H(x)$ profiles, in which waves propagate without reflection along the entire $x$-axis.
Class A contains RL flows in self-consistent channels with currents, and these flows are also regular.
One can set the profile of one of the velocities (for example, $U(x)$) on the entire $x$-axis in the form of an arbitrary continuous positive function and, using equation (\ref{M1}), obtain a family of corresponding profiles for another velocity ($c(x)$) ``labelled'' by the parameter $\Pi$.

 The class of RL flows controlled by equation (\ref{Eq-aa}) with ${\cal B} \ne 0$ (the B-class of flows) has been studied in detail by  \citet{ChSt22}. Below we consider the C-class of RL flows obeying equation (\ref{Eq-AA}) with ${\cal C} \ne 0$.
 \section{C-class RL flows}
 \label{sec3}
 \subsection{The distinctive features of C-class flows compared to the B-class flows}
 \label{sec3-1}
 \hspace\parindent
The similarities and differences in the behavior of C-class and B-class of flows are determined by the similarities and differences between equations (\ref{Eq-AA}) and (\ref{Eq-aa}). 
Equation (\ref{Eq-aa}) is homogeneous in the velocities $c$ and $U$, and the constant ${\cal B}$ has the dimension of inverse length, whereas (\ref{Eq-AA}) does not have this property, and ${\cal C}$ has the dimension of acceleration. Therefore, in the B-class of flows one can introduce the dimensionless coordinate $\xi = {\cal B}\, x$ regardless of the velocity scale $c_0$, and in the C-class, the dimensionless variables can be introduced only through the following scaling:
 \be
 \tilde{\xi} = {\cal C}x/c^2_0, \quad \tilde{c} = c/c_0, \quad
 \tilde{U} = U/c_0, \quad \tilde{a} = a/c_0.
 \label{scaleC}
 \ee
 Omitting further tilde, we rewrite (\ref{Eq-AA}) in the dimensionless form:
 \be
 \dfrac{\dd(c\,U)}{\dd\xi} =
 \dfrac{2\,c^{3/2}(\xi)U^{1/2}(\xi)}{c^2(\xi) - U^2(\xi)}\,.
 \label{Eq-sU}
 \ee
 It should be noted that due to the invariance of (\ref{Eq-AA}) with respect to the simultaneous replacement of $x \to -x$ and $ {\cal C} \to -{\cal C}$, the transition to the coordinate $\xi$ removes, to a certain extent, the distinction between the concepts of ``upstream'' and ``downstream''. 
 Indeed, if $c(x)$ and $U(x)$ satisfy equation (\ref{Eq-AA}) for ${\cal C} = {\cal C}_0$, then $c(-x)$ and $U(-x)$ satisfy the same equation for ${\cal C} = -{\cal C}_0$, but, anyway, $U(x) > 0$. 
 For this reason, further we will use the terms `to the left' (`to the right') in the sense of `in the direction of decreasing (increasing) the coordinate $\xi$'.

 The right-hand side of  (\ref{Eq-sU}) is singular for $U = c$, $U = 0$, $c = 0$, as well as for unbounded growth of $U(\xi)$ and/or $c(\xi)$. To solve the question of whether these singularities are attainable at a finite $\xi$, we rewrite equation (\ref{Eq-AA}) in the form:
 \be
 \dfrac{\dd}{\dd\xi}\,\ln a(\xi) = \dfrac{c^{1/2}(\xi)}
 {U^{1/2}(\xi)\Bl[c^2(\xi) - U^2(\xi)\Br]}\,.
 \label{Eq-La}
 \ee
 It is easy to see that if $c(\xi)$ is bounded everywhere, i.e. $0 < c(\xi) < \infty$, then  not only $U = c$ can be reached at a finite $\xi$, as in the B-class of flows, but also $U = 0$ can be reached at some other finite point $\xi$. 
 Similarly, if $U(\xi)$ is bounded everywhere, the singularity $U = \infty$, as well as $c = 0$ and $c = \infty$ are attainable only asymptotically.

  For further consideration, it is convenient to introduce functions determined by the ratio of the velocities $U(\xi)$ and $c(\xi)$ in each point $\xi$:
 \be
 F(\xi) = \left[\dfrac{U(\xi)}{c(\xi)}\right]^{1/2} \equiv
 \left[\dfrac{U^2(\xi)}{g H(\xi)}\right]^{1/4} \quad \mbox{and}
 \quad f(\xi) = \dfrac{1}{F(\xi)}\,.
 \label{Fr}
 \ee
 For the sake of brevity, we will call function $F(\xi)$ the Froude number, and function $f(\xi)$ the reciprocal Froude number (in \citep{ChSt22} they were denoted as $u(\xi)$ and $w(\xi)$). 
 In terms of $F$ and $f$, equations (\ref{Eq-sU}) and (\ref{Eq-La}) have the form:
  \bea
 \label{E-F}
 c^2(\xi)\,\dfrac{\dd F}{\dd\xi} & = & \dfrac{1}{1-F^4(\xi)} - M(\xi)F(\xi),
   \\
 \label{E-f}
 c^2(\xi)\,\dfrac{\dd f}{\dd\xi} & = & \dfrac{f^6(\xi)}{1-f^4(\xi)} +
 M(\xi)f(\xi),
   \\
 c(\xi)\,\dfrac{\dd a}{\dd\xi} & = & \dfrac{1}{1 - F^4(\xi)}\ \ = \ \
 \dfrac{f^4(\xi)}{f^4(\xi) - 1}\,,
 \label{E-a}
 \eea
 where
 \be
 M(\xi) = c(\xi)\,\dfrac{\dd c}{\dd\xi} \equiv
 \dfrac{\dd}{\dd\xi}\left(\dfrac{c^2(\xi)}{2}\right) \equiv
 \dfrac{g}{2}\,\dfrac{\dd H(\xi)}{\dd\xi}
 \label{Mu}
 \ee
 is determined by the bottom slope.
 It is convenient to present solutions of these equations as a set of trajectories (the phase portrait) on the half-plane $(\xi, \, F)$ or $(\xi, \, f)$ (recall that functions $F$ and $f$ are positive).

As an useful illustration, consider flows in channels of a constant depth, where the wave velocity is also constant, $c(\xi) = c_0$. Setting $c_0 = 1$ and integrating  (\ref{E-F}), we arrive at the algebraic equation
 \be
 F^5(\xi) - 5F(\xi) + 5(\xi - \xi_0) = 0, \qquad \xi_0 = \mbox{const}.
 \label{F5}
 \ee
 If $\xi_0 < \xi \le \xi_* = \xi_0 + 4/5$ then, this equation has two positive roots $F_\pm(\xi)$ which merge in one double root $F=1$ at $\xi = \xi_*$.
In the vicinity of the point $\xi_*$ these solutions are:
 \be
 F_\pm(\xi) \approx 1 \pm\frac{1}{2}\Bl(\xi_*-\xi\Br)^{1/2} -
 \frac{1}{4}\Bl(\xi_*-\xi\Br) + \dots
 \label{F-c}
 \ee
The bigger root, $F_+ \ge 1$, grows indefinitely when $\xi$ decreases from $\xi_*$ up to minus infinity, whereas the smaller root, $F_- \le 1$, changes its sign at $\xi = \xi_0$:
 \be
 F_-(\xi) = \xi - \xi_0 + \frac{1}{5}(\xi - \xi_0)^5 + \dots,
 \label{F-}
 \ee
 so that for $\xi < \xi_0$ equation (\ref{F5}) has only a single positive root.
 Thus, $\xi_0$ is the singular point for subcritical flow in which  $U(\xi)$ vanishes, $U(\xi) \sim (\xi-\xi_0)^2$.
 
\begin{figure}
\vspace{-0.5cm}
 \epsfxsize=120mm
\centerline{\epsfbox{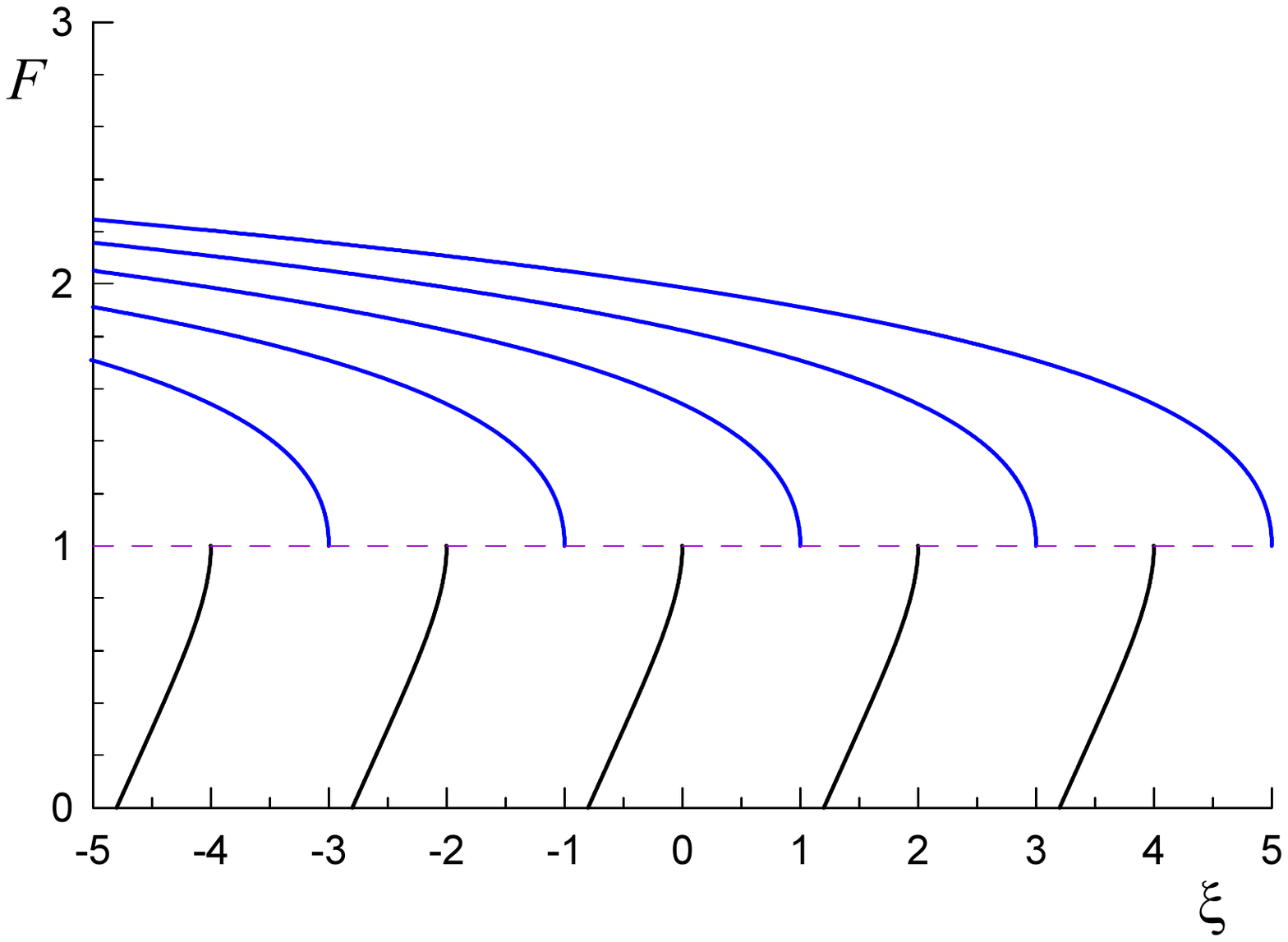}}
\vspace{-9cm}
 \caption{Phase portrait of C-class flows in a channel of a constant depth in subcritical ($F < 1$) and supercritical ($F > 1$) regions.}
 \label{fg3}
 \end{figure}

Thus, in channels of constant depth, subcritical flows of C-class (as opposed to those of B-class) remain RL only within a finite interval of  $\xi$, \ \ $\xi_0 < \xi < \xi_*$, and supercritical flows are RL on the semi-axis $\xi < \xi_*$ (see figure \ref{fg3} and compare it with the figure 3(a) in \citep{ChSt22}).
Let us find the conditions under which these restrictions are absent on some part of trajectories.
%
 \begin{figure}[!h]
 \epsfxsize=130mm
\centerline{\hspace{-2cm}\epsfbox{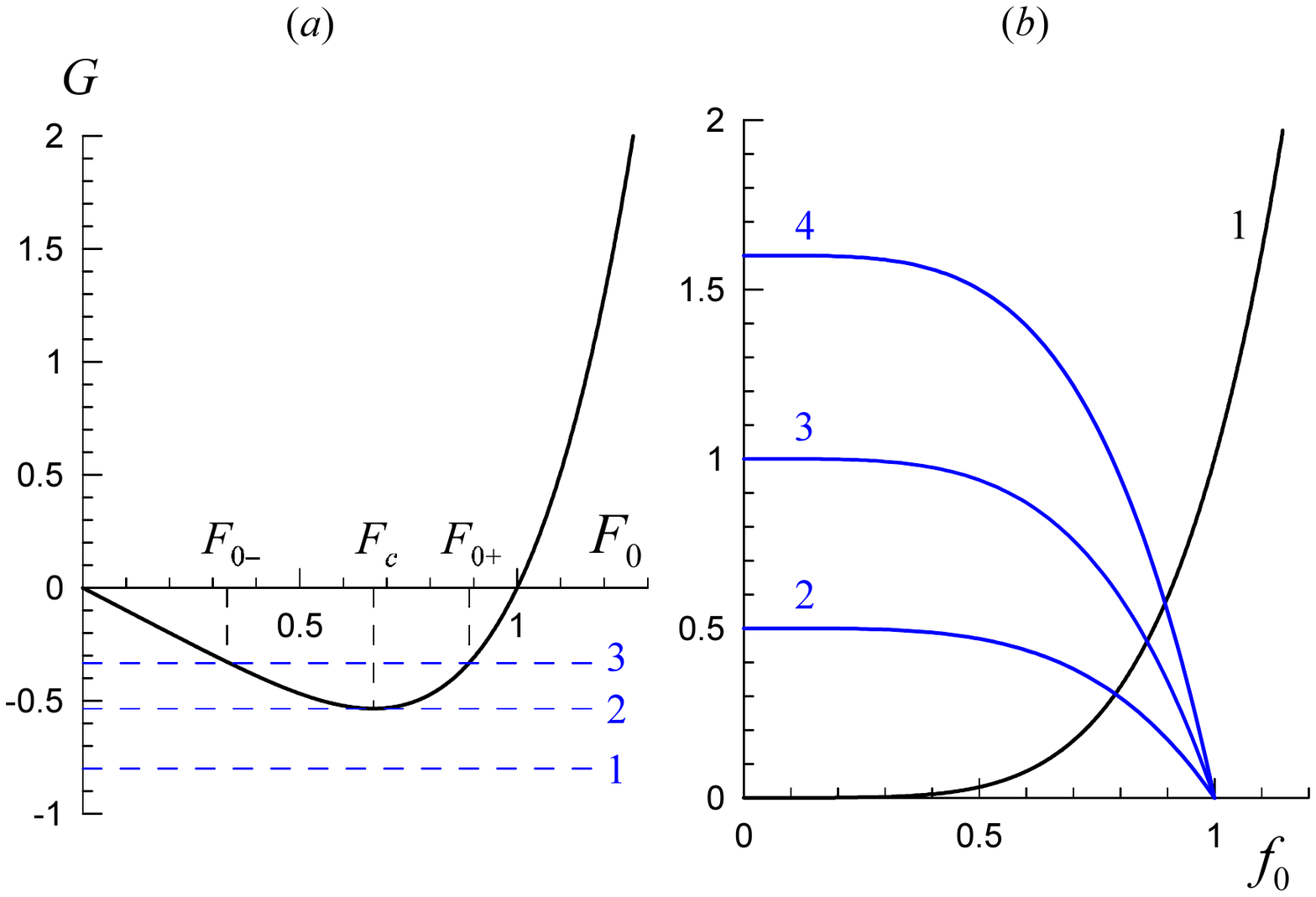}}
 \vspace{-10cm}
 \caption{Roots of equations for null-isoclines. \ (a) \  Equation (\ref{F0M}): \ dashed lines \ 1 -- $M = 1.25$, \ \  2 -- $M = M_c$, and \ \  3 -- $M = 3$. \ \ (b) \ \ Equation (\ref{Mf0}): left-hand side (curve 1) and the right-hand side at $M = -0.5$ (curve 2), $ M = -1 $ (curve 3), and $M = -1.6$ (curve 4).}
 \label{fg4}
 \end{figure}
 \subsection{Global trajectories and asymptotic behaviour}
 \label{sec3-2}
 \hspace\parindent
 A subcritical trajectory to be unbounded in $\xi$, i.e. to be {\it global}, must reach neither $F = 1$ when $\xi$ increases, nor $F = 0$ when $\xi$ decreases. 
 Thus, the task is split into two parts. Let us find first the conditions under which a trajectory is not bounded from the right. 
 As in the B-class flows, reaching the value $F = 1$ can only be prevented by the presence on the phase plane of regions with opposite signs of the right-hand side of  (\ref{E-F}), separated by the null-isocline (NI). 
 NI is described by the equation:
 \be
 G(F_0) = F^5_0(\xi) - F_0(\xi) =-M^{-1}(\xi).
 \label{F0M}
 \ee
This equation has two positive roots, $0 < F_{0-}(\xi) \le F_{0+}(\xi) < 1$ if (see figure \ref{fg4}\,(a))
 \be
 M(\xi) \ge M_c = \frac{5^{5/4}}{4} \approx 1.8692.
 \label{Mc}
 \ee
 
 Thus, in the subcritical region, NI appears only at a sufficiently large slope of the channel bottom as a result of the merger of two complex conjugate roots of  (\ref{F0M}). NI has two branches that cannot extend far to the left. Indeed, if $c(\xi_1) = c_1 > 0$ and $M(\xi) \ge M_c$ for $\xi < \xi_1$ then, with decreasing $\xi$, we will inevitably arrive at the singularity $c = 0$ ($H = 0$) for a finite $\xi$.

\begin{figure}
 \epsfxsize=130mm
\vspace{-3.0cm}
\centerline{\epsfbox{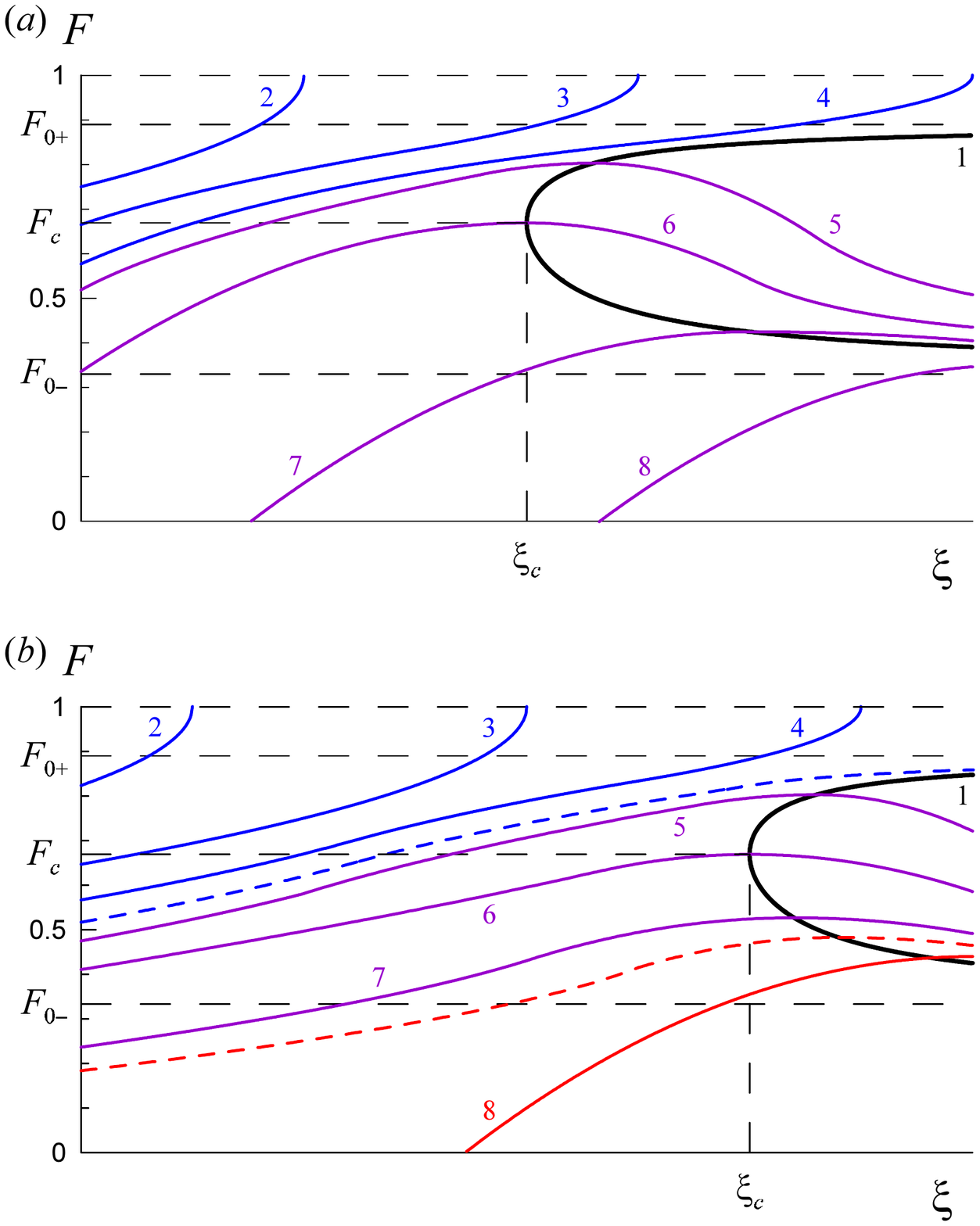}}
 \vspace{-2.0cm}
 \caption{The subcritical part of the phase portrait of equation (\ref{E-F}) for $M_0 = 3$. (a) Null-isocline (curve 1) and surrounding trajectories, bounded (curves 2--4) and unbounded (curves 5--8) on the right; trajectories 7 and 8 are bounded from the left by the singularity $F = 0$. (b) Bounded (curves 2--4 and 8) and global (curves 5--7) trajectories in the presence of a null-isocline (curve 1) and with inequality (\ref{suff}) fulfilled; blue and red dashed lines show the boundaries of the bundle of global trajectories.}
 \label{fg5}
 \end{figure}

Let us assume that $M(\xi) = M_c$ for $\xi = \xi_c$ and grows monotonically for $\xi > \xi_c$. 
Then NI branches, $F = F_{0\pm}(\xi)$, start at the point $\xi = \xi_c$ and each monotonically tends to its own limit (see figure \ref{fg5}). 
The slope of the trajectories is negative between the branches and positive outside. 
Therefore, trajectories passing above $F_{0+}(\xi)$ end on reaching $F = 1$ at finite $\xi$. 
But any trajectory that crosses any branch remains between them up to $\xi = +\infty$, i.e. is not bounded on the right, as well as all trajectories lying below it (see figure \ref{fg5}\, (a)).

 Monotonic growth of $M(\xi)$ does not require so fast an increase in depth. 
 In the boundary case, when $M(\xi)$ tends to the finite limit $M_0 > M_c$ when $\xi \to +\infty$,
 \be
 H(\xi)\sim M_0\xi, \quad F(\xi) \to F_{0-} > 0, \quad  U(\xi) \approx
 F^2_{0-}c(\xi) \sim \xi^{1/2}, \quad W(\xi) \sim \xi^{-3/2},
 \label{Fi+}
 \ee
 that is, the  flow and wave
 velocities grow in the same way, and the channel is narrowed.

As in the B-class of flows, the asymptotic (for $\xi \to \pm \infty $) behavior of subcritical flows depends on the convergence at the upper limit of the integrals
 \be
 I_{F\pm}(\xi) = \pm\din_{\xi}^{\pm\infty}\!\dfrac{\dd y}{c(y)}.
 \label{IFpm}
 \ee
 The convergence requires that function $c(\xi)$ must grow with $\xi$ faster than a linear function, for example, as $ |\xi|^{1+\vep}$, where $\vep > 0$.

  Let function $M(\xi)$ grow unlimitedly, so that $F_{0-}(\xi) \sim M^{-1}(\xi) \to 0$. 
  Consider a trajectory passing through the point $(\xi_1, \, F_1)$ into the region $\xi > \xi_1$, and denote $c_1 = c(\xi_1)$ and $a_1 = c_1F_1 $. 
  From equation (\ref{E-a}) we find:
  \be
 a(\xi) = a_1 + \din_{\xi_1}^{\xi}\dfrac{\dd y}{c(y)[1-F^4(y)]} =
 a_1 + \dfrac{1}{1-F^4(\xi_a)}\din_{\xi_1}^{\xi}\dfrac{\dd y}{c(y)},
 \label{sFC}
 \ee
 where $\xi_a$ lies between  $\xi_1$ and $\xi$. 
 If the integral $I_{F +}(\xi_1)$ converges, than $a(\xi)$ tends to the limiting value $a_{1+} > 0$, and the asymptotic relations hold (cf. (\ref{scc})):
 \be
 W(\xi) \sim U(\xi) \sim F(\xi) \sim H^{-1/2}(\xi) \sim c^{-1}(\xi).
 \label{as-A}
 \ee
 If $c(\xi)$ grows slower than $\xi$, for example, as $\xi^p$, where $1/2 \le p < 1$, than the integral $I_{F +}(\xi_1)$ diverges, and the following asymptotic relations are valid:
 \be
 a(\xi) \sim \xi^{1-p}, \quad F(\xi) \sim \xi^{1-2p}, \quad U(\xi)
 \sim \xi^{2-3p}, \quad W(\xi) \sim \xi^{p-2}, \quad H(\xi) \sim \xi^{2p}.
 \label{asC+}
 \ee
 For $p = 1/2$ these relations reduce to (\ref{Fi+}), and when $p < 1/2$, NI disappears, and all trajectories are bounded on the right.

Consider now the continuation of the trajectory passing through the point $(\xi_1, \, F_c)$ to the left, into the region $\xi < \xi_1$. For $a(\xi)$ not to vanish (together with $F(\xi)$) at some finite $\xi$, the integral $ I_{F-}(\xi_1)$ must converge, and  $\lim_{\xi \to -\infty} a(\xi) = a_{1-}$ must be positive. Since $M(\xi) < 0$ and $F(\xi)$ grows monotonically, the following inequalities hold:
 \[
 a_1 - \dfrac{I_{F-}(\xi_1)}{1-F^4_1} < a_{1-} < a_1 - I_{F-}(\xi_1).
 \]
 For the unlimited continuation of the trajectory to the left, it is necessary that
 \be
 a_1 \equiv c_1F_1 > I_{F-}(\xi_1).
 \label{neces}
 \ee
 Because $\max\Bl[F(1-F^4)\Br] = M_c^{-1}$ at $F = F_c = 5^{-1/4}$, we obtain the condition 
 \be
 c_1 \ge M_c\,I_{F-}(\xi_1),
 \label{suff-b}
 \ee
 sufficient for the trajectory passing through the point $(\xi_1, \, F_c)$ to continue with no limit to the left as well.
 Together with this trajectory, all the above-lying (with $F_1 > F_c$) and some part of below-lying (with $F_1 < F_c$) trajectories are also continue with no limit to the left. The condition (\ref{neces}) cuts off low-lying trajectories which inevitably reach $F = 0$ at some finite $\xi$ (curves 7 and 8 in figure \ref{fg5}\,(a) and curve 8 in figure \ref{fg5}\,(b)).

 Thus, we see that for the existence of global subcritical flows of C-class, the channel depth $H(\xi)$ must increase indefinitely both to the left (faster than $\xi^2$), for the inequality
 \be
 c(\xi_c) \ge M_c\,I_{F-}(\xi_c)
 \label{suff}
 \ee
 to be hold, and to the right (faster than $M_c \xi$) to ensure the monotonic growth of $M(\xi)$ for $\xi > \xi_c$, which is necessary to maintain NI. 
When these conditions are met, the set of global trajectories forms a bundle of trajectories strung on the trajectory passing through the point $(\xi_c, \, F_c)$ (in figure \ref{fg5}\,(b) the bundle boundaries are shown by dashed lines). 
Note that in the supercritical part of the phase portrait ($F > 1$), all trajectories are bounded on the right by the singularity $F = f = 1$, as in figure \ref{fg3}.

Global {\it supercritical} trajectories can arise if to the right of some point $\xi_m > -\infty$ function $c(\xi)$ decreases monotonically (i.e. $M(\xi) < 0$) that leads to the appearance of NI $f = f_0(\xi)$, described, according to (\ref{E-f}), by the equation
 \be
 f^5_0(\xi) = -M(\xi)\Bl[1-f^4_0(\xi)\Br].
 \label{Mf0}
 \ee
 As seen in figure \ref{fg4}(b), for any $M < 0$ there is one positive root $f_0 < 1$ such that
 \bea
 \label{asf-0}
 f_0(\xi) & = & \Bl[-M(\xi)\Br]^{1/5} + \frac{1}{5}\,M(\xi) + O(|M|^{9/5}),
 \quad (-M)\ll 1, \\
 \label{asf-i}
 f_0(\xi) & = & 1 + \frac{1}{4}\,M^{-1}(\xi) + O(M^{-2}), \phantom{WWWWW}
 (-M)\gg 1.
 \eea

The existence of global solutions and the asymptotic behavior of $f(\xi)$ depend on the convergence at the upper limit of integrals (cf. (\ref{IFpm}))
 \be
 I_{f\pm}(\xi) = \pm\din_{\xi}^{\pm\infty}\! c^3(y)\,\dd y .
 \label{Ifpm}
 \ee
 For the trajectory passing through the point ($\xi_2, \, f_2$), we write (\ref{E-a}) in the form:
 \[
 c(\xi)\,\dfrac{\dd a}{\dd\xi} = -\dfrac{f^4(\xi)}{1-f^4(\xi)} =
 -\dfrac{c^4(\xi)}{a^4(\xi)[1-f^4(\xi)]}.
 \]
and integrate it:
 \be
 a^5(\xi) \equiv \left[\dfrac{c(\xi)}{f(\xi)}\right]^5 = a^5(\xi_2) -
 5\!\din_{\xi_2}^{\xi}\dfrac{c^3(y)\,\dd y}{1-f^4(y)} =
 \left[\dfrac{c(\xi_2)}{f_2}\right]^5 -
 \dfrac{5}{1-f^4(\xi_b)}\din_{\xi_2}^{\xi}\!c^3(y)\,\dd y,
 \label{sfC}
 \ee
where the point $\xi_b$ lies between $\xi_2$ and $\xi$. The trajectory will be global if the integral $I_{f+}(\xi_2)$ converges and $f_2$ is small enough for the positiveness of the limit $a^5_{2+}$ of the right-hand side of equation (\ref{sfC}) when $\xi \to +\infty$. Then $a(\xi) \to a_{2+}$, and relations (\ref{as-A}) are valid.

For $\xi \to -\infty$, all trajectories are unlimited, but the behavior of function $f(\xi)$ depends on the convergence of the integral $I_{f-}(x_2)$. 
If it converges, $a(\xi) \to a_{2-} > 0$, and relations (\ref{as-A}) hold. 
If it diverges and, for example, $c(\xi) \sim (-\xi)^q$, where $-1/3 < q < 1/2$, then
 $$
 a(\xi) \sim (-\xi)^{(1+3q)/5}, \quad f(\xi) \sim (-\xi)^{(2q-1)/5},
 $$
 \be
 U(\xi) \sim (-\xi)^{(2+q)/5}, \quad W(\xi) \sim (-\xi)^{-(2+11q)/5}.
 \label{asC-}
 \ee
 For $q > 0$, we have $M(\xi) \sim -q(-\xi)^{2q-1} < 0$, and NI $f = f_0(\xi) \approx M^{1/5}(\xi)$ appears to which $f(\xi)$ tends asymptotically. And when $q \ge 1/2$, they both tend to a finite nonzero limit, so that in this case for $\xi \to -\infty$, we have:
 \be
 a(\xi) \sim U(\xi) \sim c(\xi), \quad W(\xi) \sim c^{-3}(\xi),
 \quad H(\xi) \sim c^2(\xi).
 \label{asC-2}
 \ee

\begin{figure}
 \epsfxsize=115mm
 \vspace{-1.5cm}
\centerline{\epsfbox{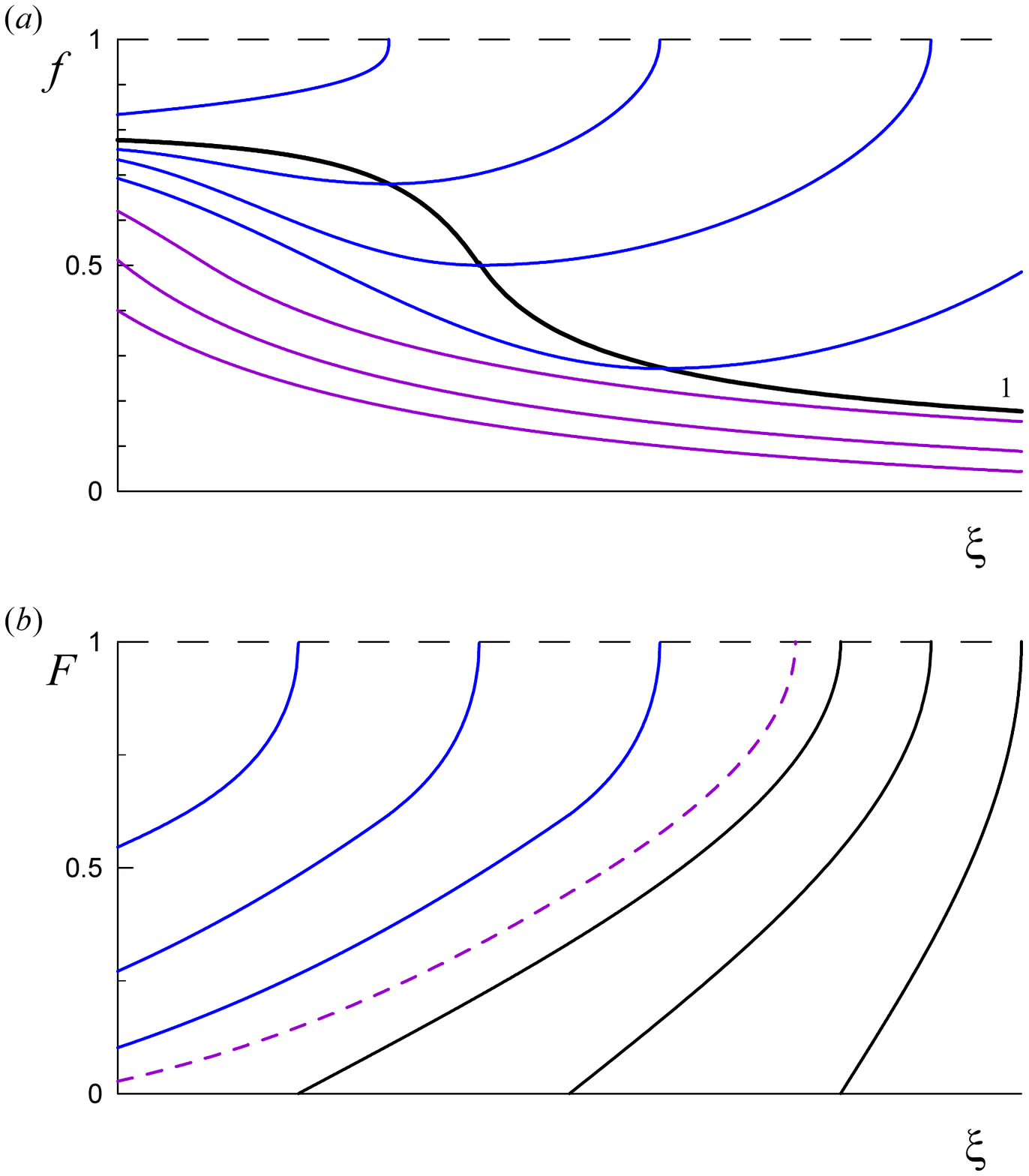}}
 \vspace{-3.5cm}
 \caption{Qualitative view of the phase portrait for $M(\xi) < 0$: (a) the supercritical part ($f < 1$, $f_0 (M_-) = 0.8$), line 1 is the null-isocline; (b) the subcritical part ($F < 1$), the dashed line shows the separatrix.}
 \label{fg6}
 \end{figure}

Let us describe in more detail the phase portrait of flows in a channel with the depth decreasing in such a manner that $M(\xi) < 0$ monotonically increases. 
Let $M(\xi)$ has a negative (finite or infinite) limit $M_-$ when $\xi \to -\infty$, whereas $M(\xi)$ goes to zero when $\xi \to +\infty$ faster than $-\xi^{-5/3}$ to secure the existence of global trajectories. 
Then NI $f_0(\xi)$ decreases monotonically from $f_0(M_-)$ (see figure \ref{fg4}\,(b)) to zero. Each trajectory lying above NI or intersecting it is bounded on the right, but there are also global trajectories that lie entirely below NI and approach it from below as $\xi \to -\infty$ (see figure \ref{fg6}(a)).

Since $M(\xi) < 0$, in the subcritical part of the phase portrait ($f > 1$, $F < 1$) all trajectories are bounded on the right by the singularity $F = f = 1$. 
Suppose, however, that for some $\xi_1$ condition (\ref{suff-b}) is satisfied, so that there are trajectories that are unbounded from the left. 
They are separated from the underlying trajectories bounded on both sides by a separatrix. For greater clarity, this part of the phase portrait is shown in figure \ref{fg6}(b) in coordinates $(\xi, \, F)$.
 \section{Concluding remarks}
 \label{sec4}
 \hspace\parindent
The C-class of RL flows considered here is in many ways similar to the B-class studied in \citep{ChSt22}. 
Indeed, in both these classes flows can be either subcritical or supercritical, since, unlike flows of A-class, the profiles $c(\xi)$ and (or) $U(\xi)$ are inevitably singular in the critical point $U = c$. Flows passing through this point are, apparently, not reflectionless.
Further, supercritical flows of both classes are not bounded on the left in $\xi$, and their asymptotic behavior and the existence of global flows are equally dependent on the behavior of $c(\xi)$,
 namely, on the convergence of integrals (\ref{Ifpm}).

 The differences, and very significant ones, show subcritical currents.
 First of all, subcritical trajectories of the C-class can be bounded in $\xi$ not only from the right (by the critical point $U = c$), but also from the left by the singularity $U = 0$ (see figure \ref{fg3} and compare with figure 3(a) in \citep{ChSt22}).
 Further, the continuation of the trajectories to the right in both classes is possible only when a null-isocline appears in the phase portrait. 
 But the C-class differs in both the geometry of NI (cf. NIs in figure \ref{fg5} and figure 7 in \citep{ChSt22}), and the need to exceed the threshold value (\ref{Mc}) of the bottom slope for its appearance. 
 Due to the $U$-shaped NI, there is no need for the convergence of the integral $I_{F+}(\xi)$ (see equation (\ref{IFpm})) for the existence of trajectories that are not bounded on the right. 
 Therefore, they appear with a slower increase in the depth of the channel $H(\xi)$ than in the B-class, and differ in a variety of asymptotic behavior, cf. equations (\ref{Fi+}), (\ref{asC+}) and (\ref{as-A}).

 A separate question that does not arise in the B-class but is important in the C-class is the continuation of subcritical trajectories to the left. 
 For this, the convergence of the integral $I_{F-}(\xi)$ is not enough, and the more stringent inequalities (\ref{neces}) or (\ref{suff-b}) must hold. 
 The simultaneous observance of the conditions for the unbounded continuation of trajectory both to the right and to the left leads to the fact that in the C-class of flows, global trajectories form a bundle bounded by the Froude number $F$ both above and below (see figure \ref{fg5}\,(b)), whereas in the B-class there is only an upper constraint (see figure 7 in \citep{ChSt22}).\\
 
{\bf Funding.} S.C. was financially supported by the Ministry of Science and Higher Education of the Russian Federation.
 Y.S. acknowledges the funding of this study provided by the grant No. FSWE-2020-0007 through the State task program in the sphere of scientific activity of the Ministry of Science and Higher Education of the Russian Federation, and the grant No. Grant No. NSH-70.2022.1.5 provided by the President of Russian Federation for the State support of leading Scientific Schools of the Russian Federation. 
 
{\bf Declaration of interests.} The authors report no conflict of interest.


{\bf Author contributions.} S.C. derived the theory. All authors contributed equally to analysing data, reaching conclusions, and in writing the paper.

{\bf Author ORCIDs.} \\
Semyon M. Churilov {\color{blue} https://orcid.org/0000-0001-5543-2474;}\\
Yury A. Stepanyants {\color{blue} https://orcid.org/0000-0003-4546-0310.}

\bibliographystyle{jfm}
\bibliography{Chur-Step-Short}

\end{document}